\documentclass[fleqn,twoside,]{article}
\usepackage{espcrc2}
\usepackage[figuresright]{rotating}
\usepackage {dcolumn}
\usepackage {graphicx}

\title{Formation of magnetic moments in the cuprate
superconductor Hg$_{0.8}$Cu$_{0.2}$Ba$_2$Ca$_2$Cu$_3$O$_{8+\delta}$
below $T_c$ seen by NQR}

\author{H. Breitzke\address[MC]{Fachbereich Physik, Freie Universit\"at Berlin,
D-14195 Berlin, Germany}, I. Eremin\address[tp]{Institut f\"ur
Theoretische Physik, Freie Universit\"at Berlin, D-14195 Berlin,
Germany}\address[kazan]{Physics Department, Kazan State
University, 420008 Kazan, Russia}, D. Manske\addressmark[tp], E.V.
Antipov\address{Department of Chemistry, Moscow State University,
119899 Moscow, Russia} and K. L\"uders\addressmark[MC]}
\date{\today}

\begin{document}
\begin{abstract}
We report pure zero field nuclear magnetic resonance (NQR)
measurements on the optimally doped three layer high-$ T_{c}
$-compounds HgBaCaCuO and HgBaCaCuO(F) with $T_c$ 134~K. Above $
T_{c}$ two Cu NQR line pairs are observed in the spectra
corresponding to the two inequivalent Cu lattice sites. Below $
T_{c}$ the Cu NQR spectra show additional lines leading to the
extreme broadened Cu NQR spectra at 4.2~K well known for the
HgBaCaCuO compounds.  The spin-lattice relaxation curves follow a
triple exponential function with coefficients depend onto the
saturation time (number of saturation pulses), whereas the
spin-spin relaxation curve is described by a single exponential
function. From the spin-lattice relaxation we deduced a complete
removal of the Kramers degeneracy of the Cu quadrupole indicating
that the additional lines are due to a Zeemann splitting of the
$^{63/65}$Cu lines due to the spontaneous formation of magnetic
moments within the CuO layers. Below 140~K, the spectra are well
fitted by a number of 6 $^{63/65}$Cu line pairs. From the number
of the Cu lines, the position of the lines relative to each other
and the complete removal of the Kramers degeneracy we deduced an
orientation of the magnetic moments parallel to the symmetry axis
of the electric field gradient tensor with magnitudes of the order
of 1000~G. We also discuss the possible microscopic origin of the 
observed internal magnetic fields.
\end{abstract}
 \maketitle

\section{Introduction}

A possible co-existence of antiferromagnetism and
superconductivity remains one of the most interesting problem in
high-$T_c$ cuprates. Despite that in several theoretical scenarios
of high-$T_c$ superconductivity in cuprates the Cooper-pairing
arises due to an exchange of antiferromagnetic spin fluctuations
present in the paramagnetic phase \cite{chubuk,manske}, it is
commonly believed that bulk superconductivity and the
antiferromagnetism do not coexist in the cuprates at the same
doping concentration and temperature. However, recently possible
formation of antiferromagnetism below the superconducting
transition temperature was found by several experimental
techniques in underdoped YBa$_2$Cu$_3$O$_x$ and
La$_{2-x}$Sr$_x$CuO$_4$ \cite{mook,sidis,sonier,lake}. The
relatively small values of the observed magnetic moments
\cite{mook,sidis} ($\sim 0.01~\mu_B$, where $\mu_B$ is the Bohr
magneton) have indicated an orbital rather than an electronic
origin of the observed antiferromagnetism. Interestingly,
some of these findings were interpreted in terms of the
time-reversal symmetry breaking state\cite{varma} or $id$-density
wave state \cite{laughlin} that may occur in the underdoped
cuprates below the pseudogap formation temperature, $T^*$. Most
recently, the time-reversal symmetry breaking below $T^*$ has been
indeed observed in underdoped Bi$_2$Sr$_2$CaCu$_2$O$_{8+\delta}$
by means of the angle-resolved-photoemission spectroscopy (ARPES)
\cite{kaminski}.

\begin{figure}[t]
\vspace*{-1.0cm}
\includegraphics[width=6cm]{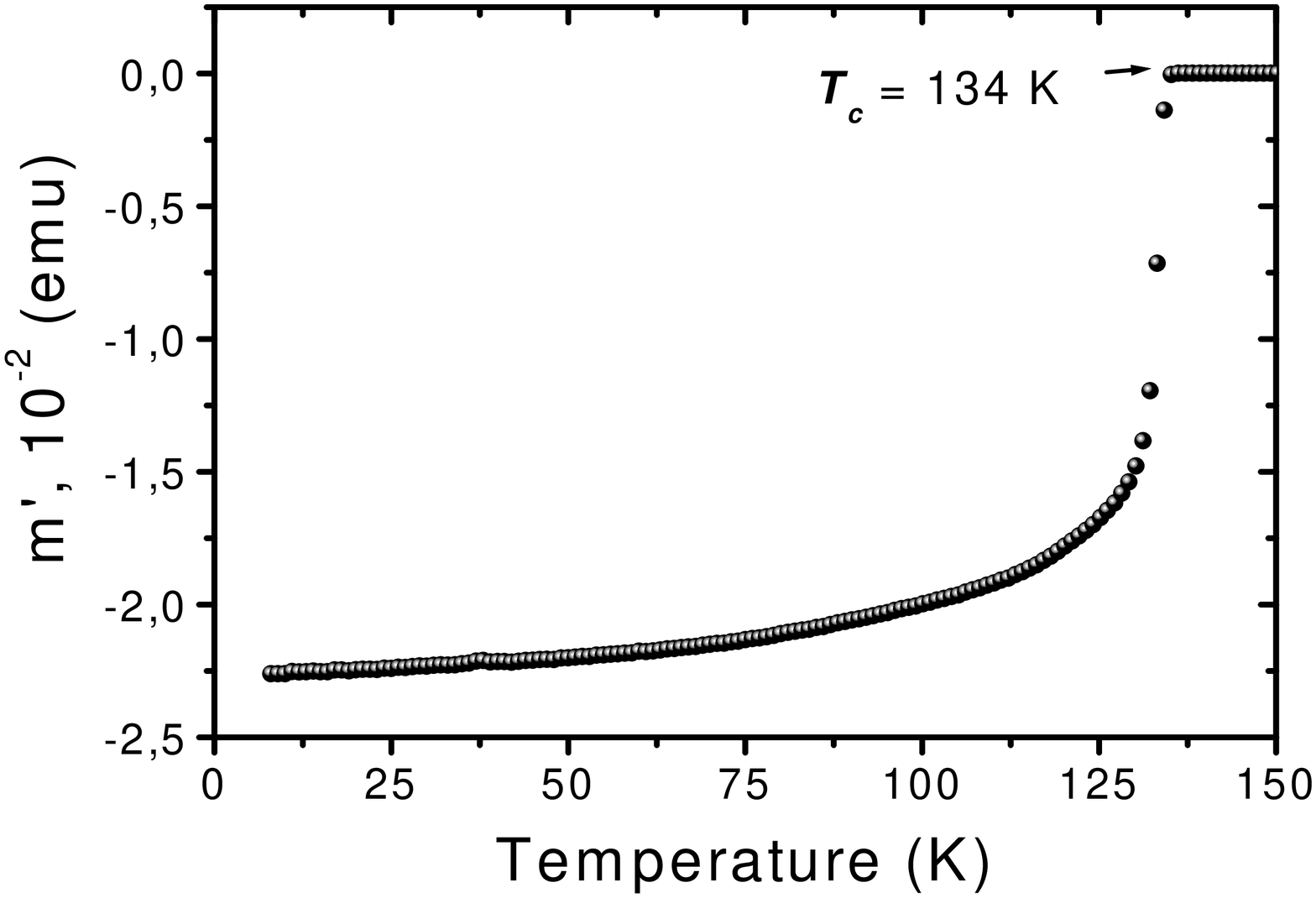}
\includegraphics[width=6cm]{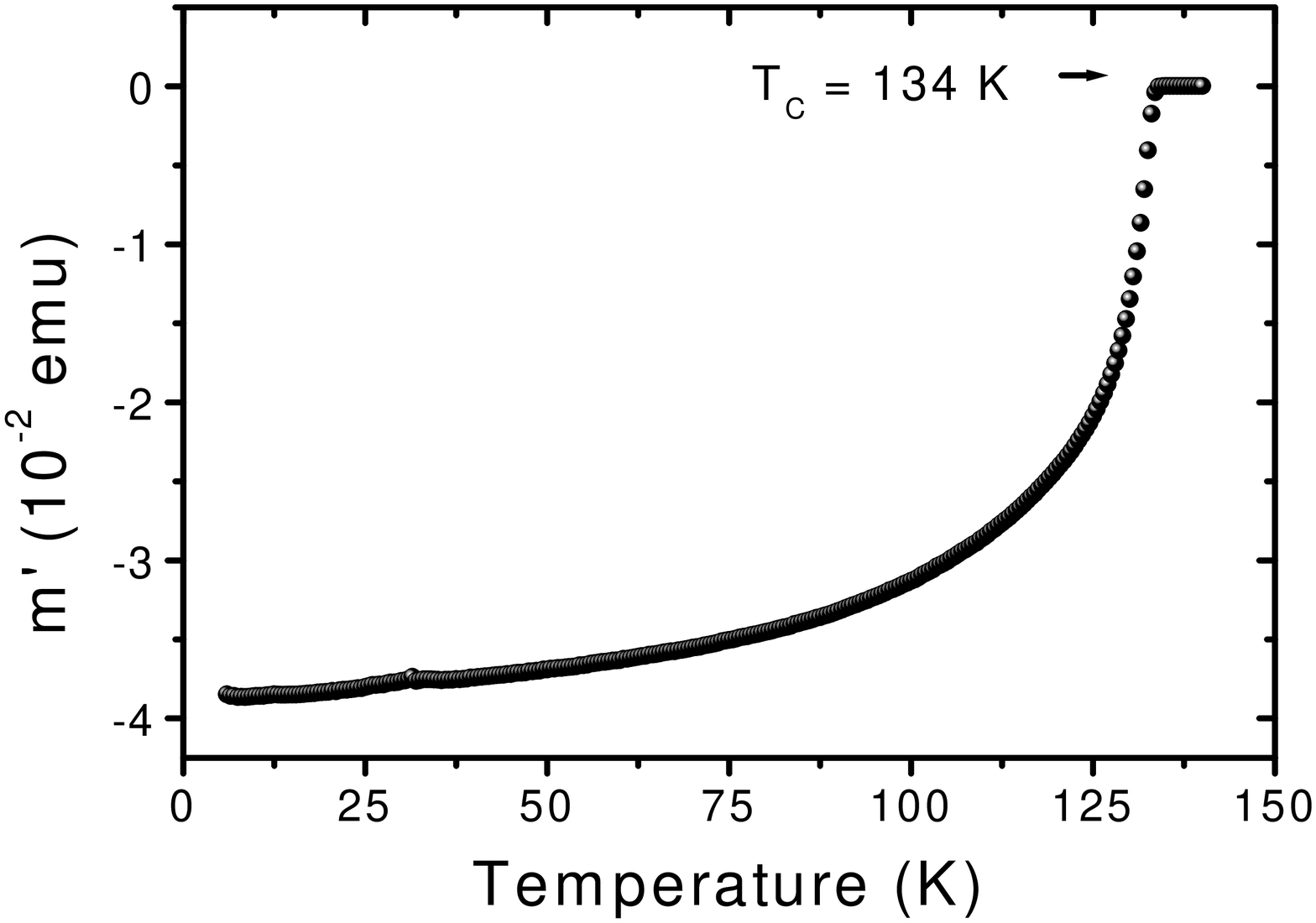}
\caption{\label{suscep}Real part of the ac-susceptibility of the
oxiginated sample (upper picture) of the fluorinated sample (lower
picture). The superconducting transitions are relatively sharp and
shows no further phases. }
\end{figure}
Nuclear magnetic resonance (NMR) and nuclear quadrupole resonance
(NQR) can be one of the most powerful tools to study the possible
occurrence of the intrinsic antiferromagnetism in high-$T_c$
cuprate superconductors. Very recently the maximum in the
transverse relaxation rate of Cu(2) nuclei observed in
YBa$_2$Cu$_3$O$_7$ below $T_c$ at T~=~35~K was interpreted in
terms of critical fluctuations of the $id$-density wave state
\cite{eremin}. A similar conclusion was drawn from the analysis of
the spin-freezing processes in underdoped
Y$_{2-x}$Ca$_x$Ba$_2$Cu$_3$O$_6$ and
La$_{2-x}$Sr$_x$CuO$_4$\cite{riga}. On the other hand, the NQR
study provides a more direct prove for the formation of orbital
magnetism since it is performed in zero magnetic field. Thus, the
internal magnetic moments if they are present will result in the
NQR line splitting. So far such a splitting was not observed in
high-$T_c$ cuprates. Instead, one mainly observes an anomalous
broadened Cu NQR spectra at low doping
\cite{kohori,fujiwara1,fujiwara2,oashi,lipinski,asayama}.
Associating the broad Cu NQR spectra with doping inhomogeneities
is leading to a few contradictions. The spectral width should
depend on the preparation conditions and should have strongly
decreased with improved preparation techniques developed in the
past decade. Especially for the case of Hg-cuprates, that possess
extremely broad NQR spectra, a dependency of the width of the NQR
spectra on preparation conditions has never been reported. The
second and more relevant reason against a doped inhomogeneity as
the origin of the broad Cu NQR spectra is that it will spread the
superconducting transition. An inhomogeneously distributed
electric field gradient (EFG) is directly associated to an
inhomogeneously distributed charge density in the CuO layers. The
charge density again is associated to the superconducting
transition temperature. Thus, a distribution of spheres with
distinct EFGs inside a crystal has to be associated to a
distribution of \( T_{c} \)'s. If one estimates the distributions
of \( T_{c} \)'s of HgBaCuO from the phenomenological model by
Gippius et al. \cite{gippius}, the superconducting transition
should be broadened by several tenth of K, what is clearly not the
case.

The most interesting reason for additional NQR lines is the
removing of the Kramers degeneracy of the Cu quadrupole levels due
to a formation of magnetic fields at the Cu sites. Such magnetic
fields may arise from an antiferromagnetic state formatting
spontaneously below $T_c$.

Additional lines may also occur from impurities and Cu outside the
CuO layers. However, lines from impurities should be visible at
any temperature. Furthermore, the relaxation times of Cu inside
the CuO layers are extremely short, in comparison with metallic
Cu, and driven by antiferromagnetic  fluctuations (AF) \cite{suh}
inside the CuO layers, strongly decreasing with distance. If a
line from an impurity becomes visible at a certain temperature it
can be easily identified by their distinct relaxation times with
respect to Cu inside CuO layers.

In this work using NQR technique we present
the first experimental evidence for the formation of the internal
magnetic moments in the optimally doped three-CuO$_2$-layer
Hg$_{0.8}$Cu$_{0.2}$Ba$_2$Ca$_2$Cu$_3$O$_{8+\delta}$ (Hg-1223) and
Hg$_{0.8}$Cu$_{0.2}$Ba$_2$Ca$_2$Cu$_3$O$_{8+\delta}$$F$
high-temperature cuprate superconductors (HTSC) below $T_c$ ($T_c$
= 134~K). The synthesis of the samples is described in
Ref.~\cite{lokshin}. An impurity content of less than 5~\% within
the sample was found from neutron powder diffraction and x-ray
powder diffraction~\cite{lokshin}. The real part of the
ac-susceptibility is shown in Fig~\ref{suscep}. Both samples
possess different ways of preparation and types of dopands. Thus,
any common properties have to attributed to intrinsic properties
of the superconducting state.
 In particular, we show that $^{63}$Cu NQR-lines split
below $T_c$ due to a Zeeman splitting originating from the
internal magnetic fields within the CuO$_2$-layers.

Our measurements were carried out on a home-built NQR
spectrometer. The spectra have been recorded by point-by-point
technique with a Hahn spin-echo pulse sequence with pulse lengths
of 4~$\mu$s and 8~$\mu$s. The recovery of the longitudinal
magnetization was measured by the method of saturation recovery.
Various numbers of saturating pulses (from 3 to 101) have been
used to study the dependency of the spin-lattice relaxation on the
initial conditions. The decay of the transverse magnetization was
measured by varying the pulse separation ($\tau$) of the spin-echo
sequence. In order to detect lines with strongly different $T_2$
times the spectra were scanned with first increasing pulse
separation time $\tau_2$ of the Hahn echo sequence. To detect
lines with strongly different $T_1$ times the point to be measured
was saturating by a saturating comp and after a time $\tau_1$
measured by the Hahn spin-echo sequence.

\section{Zero field NQR}

Nuclei with a spin $ I\geq 1 $ possess a quadrupole moment and
interacting with an EFG. The EFG is described by the traceless
tensor \( V_{ij}=\frac{\delta ^{2}V}{\delta x_{i}\delta x_{j}} \).
Since the EFG tensor is a traceless tensor, only the values \(
V_{zz} \) and \( V_{xx}-V_{yy} \) are needed to characterize the
EFG tensor. The EFG tensor is commonly described by the values \(
eq \) and \( \eta \) defined by the equations \( eq=V_{zz} \) and
\( \eta =\frac{V_{xx}-V_{yy}}{V_{zz}} \). In high-\( T_{c} \)
superconductors the elongated CuO octahedrons leading to the case
of axial symmetry of the EFG at the Cu site, giving \( \eta =0 \).

With the restriction to the case of axial symmetry of the EFG
tensor, the interaction energy is defined by the well known
Hamiltonian
 \begin{equation}\label{zeemann ham}
  H=\frac{e^{2}qQ}{4I(2I-1)}(3I_{z}^{2}-I^{2})
 \end{equation}
with eQ  the quadrupole moment of the nucleus. The Eigenvalues of
this unperturbed Hamiltonian are
\begin{equation}
E=\frac{e^{2}qQ}{4I(2I-1)}(3m^{2}-I(I+1).
\end{equation}
In general, there is a twofold degeneracy of the $ \pm m $
eigenvalues corresponding to the symmetry of the nucleus charge
distribution and of the EFG. For half integer spins a deviation
from the EFGs rotational symmetry, $ \eta \neq 0 $, will not
remove the degeneracy corresponding to Kramers theorem that states
that the degeneracy of half integer spins can only completely be
removed by a magnetic field.

For the case of nuclei like Cu with spin $ I=3/2 $ the set of
levels form a virtual two level system due to the Kramers
degeneracy and only one transition will be observed. Note, that
for this particular case, an observed Cu NQR line splitting can
only be due to mesoscopic charge segregation and/or the appearance
of a magnetic field at the nucleus site.

\subsection{Zeeman perturbed NQR}

The twofold degeneracy of the quadrupole level is corresponding to
the fact, that turning the nuclei end for end will leave the
electrostatic energy unchanged. Applying a magnetic field at the
nuclei will lower the symmetry of the system and lifting the
degeneracy. In general the situation is described by adding a
Zeeman term $ H_{z}=\gamma _{n}B_{0}hI_{z} $ to the unperturbed
Hamiltonian\cite{slichter} (\ref{zeemann ham})
\begin{equation}
H=\frac{e^{2}qQ}{4I(2I-1)}(3I_{z}^{2}-I^{2})+\gamma
_{n}B_{0}hI_{z}
\end{equation}
and treating the problem in first order perturbation theory. The
solutions can be divided into three cases: The magnetic field $
B_{0} $ oriented in parallel, second perpendicular to the symmetry
axis of the EFG, and third the magnetic field oriented with angles
$\neq 0^\circ, 90^\circ $ with respect to the symmetry axis of the
EFG. In the cases $ B_{0} $ parallel to the symmetry axis, the
mixing of the $ \left| \pm 1/2\right\rangle $ level vanishes and
only two lines are observed. In the case $ B_{0} $ perpendicular
to the symmetry axis the splitting of the $ \left| \pm
3/2\right\rangle $ vanishes in first order and again two lines are
observed. For case three the quadrupole levels split into four
levels with a mixing of the $ \left| \pm 1/2\right\rangle $ levels
and four lines are observed.  The three different cases are
corresponding to different orientations of the magnetic field
direction with respect to the symmetry axis of the EFG and can be
distinguished by investigating the spin-lattice relaxation
behavior under various initial conditions.

\subsection{\label{spin lattice} Spin-lattice Relaxation}
In the simplest case, of a two level system, or a virtual two
level system (degenerated system), the concept of spin temperature
can be applied to the problem and  the establishing of equilibrium
magnetization $ m(\infty ) $ is described by the single
exponential function $ m(t)=m(\infty )(1-\exp(t/\tau )) $. Thus, a
line splitting in coincidence with a single exponential
spin-lattice recovery function characterizes a line splitting  due
to a mesoscopic charge segregation and the absence of a magnetic
field. The Kramers degeneracy is left unchanged and every line in
the spectrum corresponds to a transition between degenerated $
m=\pm 3/2 $ levels. In case of unequally spaced levels (applied
magnetic field) the system will come into internal equilibrium and
into equilibrium with the lattice with distinct time constants.
Furthermore, the recovery is not exponential and depend on the
initial conditions i.e. the saturating time. In general, the
recovery of the central transition of a four level system into
equilibrium is described by a linear combination of three
exponential functions
\begin{eqnarray}
 \frac{M(\infty)-M(t)}{M(\infty)}=\sum_{i=1}^{3} a_i\exp(-W_it)
\end{eqnarray}
whereas the $a_i$ depend on the relaxation mechanism and
saturating time.
 In order to extract the correct recovery law one has to know
the dominant relaxation mechanism. In some particular cases the
recovery law for spin-lattice relaxation can be given and are well
known in magnetic resonance. For quadrupole splitting and magnetic
relaxation for example a $I=3/2$ nuclei like Cu the recovery of
the central transition after short saturation time is described by
the recovery law

\begin{eqnarray}\label{satshort}
  \frac{M(\infty)-M(t)}{M(\infty)} &=& 0.1\exp(-2Wt)\\
  &+& 0.9\exp(-12wt) \nonumber
\end{eqnarray}
 and
\begin{eqnarray}\label{satlong}
  \frac{M(\infty)-M(t)}{M(\infty)} &=& 0.4\exp(-2Wt)\\
  &+& 0.6\exp(-12wt)\nonumber
\end{eqnarray}
for long saturating time \cite{andrew,narath}. If magnetic and
quadrupole relaxation are allowed no analytic solution can be
given. The situation in NQR is slightly different since $ \left|
\pm 1/2\right\rangle $ $\longleftrightarrow$ $ \left| \pm
3/2\right\rangle $, are excited, however, the general facts hold
true that the spin-lattice relaxation is described by a multi
exponential recovery law and that the component of the recovery
\begin{figure}[t]
\includegraphics[width=6.5cm,angle=0] {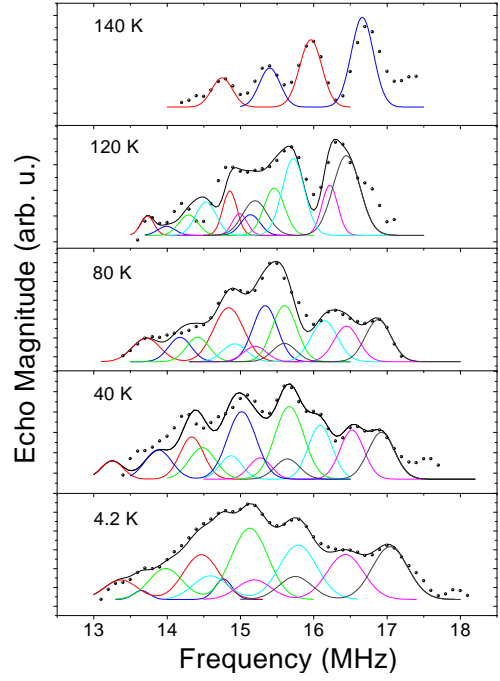}
\caption{Measured $^{63/65}$Cu NQR spectra for the oxygenated
sample at various temperatures (circles). Below $T_c$ the spectra
are described by six pairs of lines (solid curves) indicating the
splitting of the resonance frequencies. For comparison also
normal-state data are displayed where only two pairs of lines are
present.} \label{fig1}
\end{figure}
law with the longer relaxation time will increase with increasing
saturating time. This characteristic behavior can be utilized to
distinguish not only between a magnetic and electric line
splitting. In case of a magnetic line splitting it can be utilized
to distinguish between the various orientations of the magnetic to
the symmetry axis of the EFG tensor.

In case of the orientation of the magnetic field perpendicular to
EFG's symmetry axis the $ \left| \pm 3/2\right\rangle $ levels
coincidence and the system consist of three levels. The relaxation
is now described by two transitions between this three levels and
thus, in general, described by a two exponential recovery law. In
the two other cases the recovery law is, in general, described by
a three exponential law.

In most cases the time constants of the recovery law are close to
each other and the three distinct exponential functions can hardly
be resolved. In such a case one can not distinguished  a three
exponential recovery law from a two exponential recovery law by a
fitting procedure. However, three and two exponential recovery
laws can be distinguish by their answers to various saturating
times. In case of a two exponential recovery law varying the
saturating time will only affect the coefficients of the
individual exponential functions and not the time constants. If,
however, a three exponential recovery trace is fitted with a two
exponential recovery law, varying the saturating time will obtain
different time constants for various saturating times.

Thus we safely conclude, that we can distinguish a line splitting
due to a mesoscopic charge segregation and due to a removal of the
Kramers degeneracy, by recording the spin-lattice relaxation curve
under different initial conditions. Furthermore, in case of a
removal of the Kramers degeneracy the orientation of the magnetic
with respect to the symmetry axis of the EFG can be deduced from
the spin-lattice relaxation curve. Different initial conditions
may established e. g. by using the method of saturation recovery
and applying different numbers of saturation pulses. Under the
condition of mesoscopic charge segregation, one expects a single
exponential recovery independent of the number of saturating
pulses. However, a removal of the Kramers degeneracy is leading to
a multi exponential recovery curve depending on the initial
conditions.

\section{Results}

\subsection{Cu NQR Spectra}

In Fig.~\ref{fig1} we show the evolution of $^{63/65}$Cu spectra
with temperatures from T~=~140~K to T~=~4.2~K for the originated
sample. The corresponding spectra for the fluorinated sample are
\begin{figure}[t]
\includegraphics [width=6.5cm, angle=0] {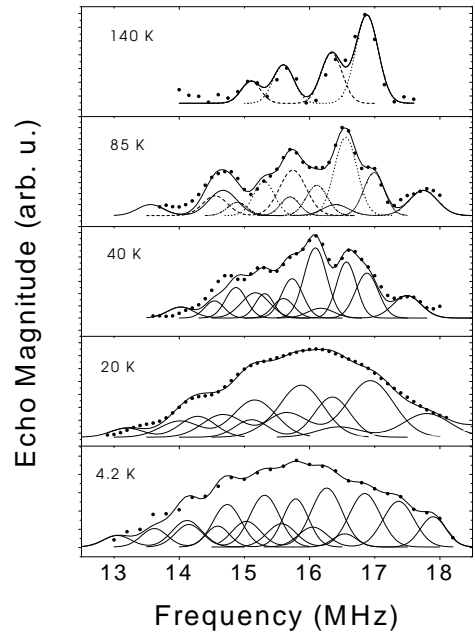}
\caption{\label{spectraf9}Measured $^{63/65}$Cu NQR spectra for
the fluorinated sample at various temperatures (circles). Below
$T_c$ and above 4.2 K the spectra are described by six pairs of
lines (solid curves). At 4.2 K the spectrum can only satisfactory
be fitted with 8 pairs of lines.}
\end{figure}
shown in Fig.~\ref{spectraf9}. For all temperatures the spectra
lie between 13~MHz and 19~MHz. Above $T_c$ (T~=~140~K) one clearly
sees two pairs of lines corresponding to the resonance transition
between $|\pm 3/2> \to |\pm 1/2>$ quadrupole levels of the copper
nuclear spins belonging to the inner and outer CuO$_2$-layers. The
spectra above $T_c$ are fitted with equation~(\ref{gausspaar})
according to the various natural abundance and quadrupole moments
of the $^{63/65}$Cu isotopes.
\begin{eqnarray} \label{gausspaar} Y&=&y_{0}+A\cdot
\exp\frac{(x-x_{0})^{2}}{2W^{2}}\\
    &+&2,3\cdot A\cdot
\exp\frac{(x-1,082\cdot x_{0})^{2}}{2\cdot (1,082\cdot
W)^{2}}\nonumber.
\end{eqnarray}
In accordance with previous NMR studies the pair of lines with
higher frequency and larger intensity corresponds to the two outer
CuO$_2$-layers while the other pair of lines to the inner
one~\cite{julien}. Most importantly, the $^{63/65}$Cu NQR of both
sample exhibit the same evolution with temperature.

Below $T_c$ the copper NQR spectra are strongly broadened and can
not be fitted with two pairs of lines. Instead, a good agreement
can be achieved by using six pairs of lines according to
equation~(\ref{gausspaar}). It is important to note that the
spectra are fitted with line pairs of fixed relative frequency
spacings and relative intensities. This restricts the degree of
\begin{figure}[t]
\includegraphics [width=6cm,angle=0]{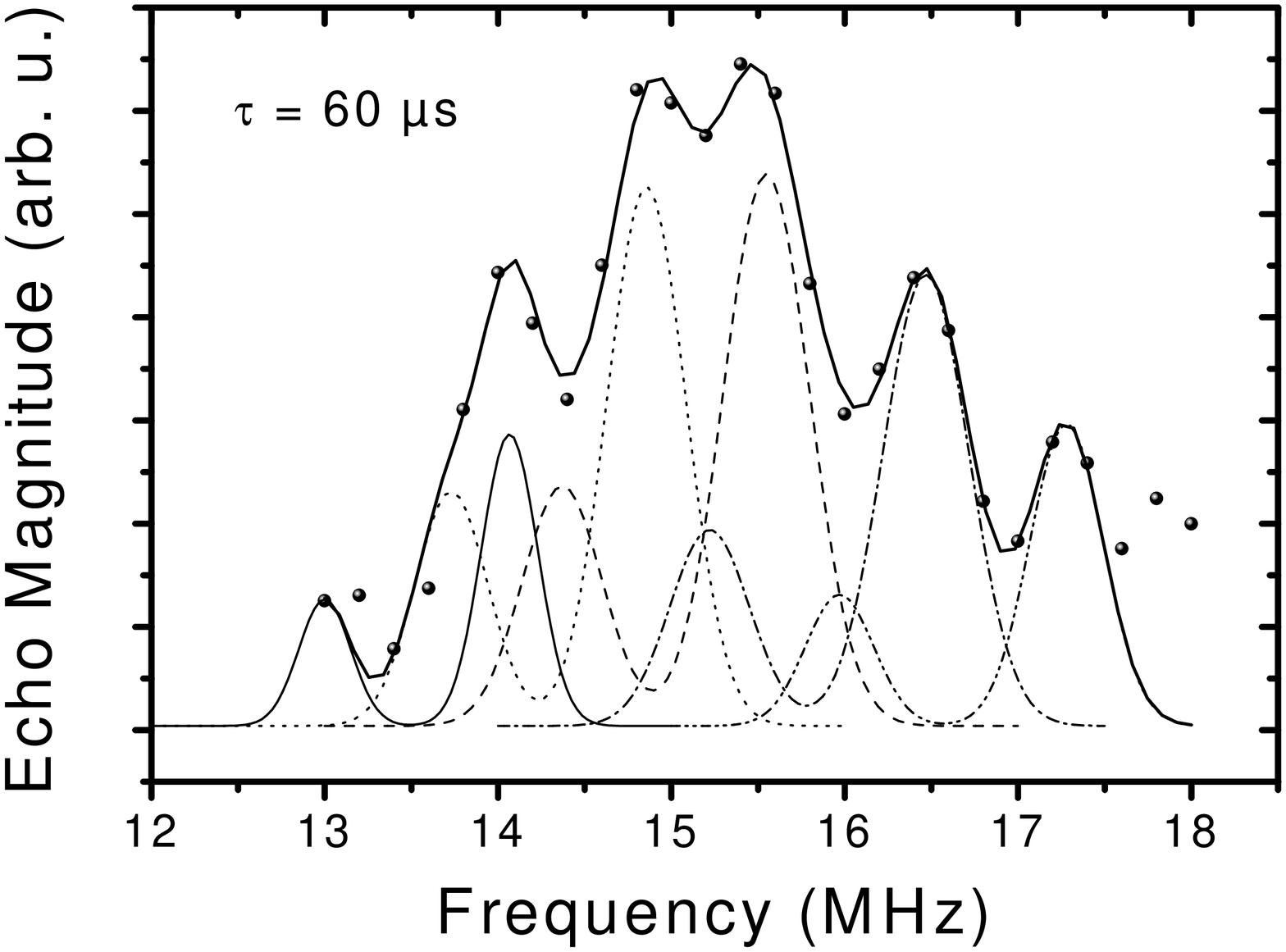}
\includegraphics [width=6cm,angle=0]{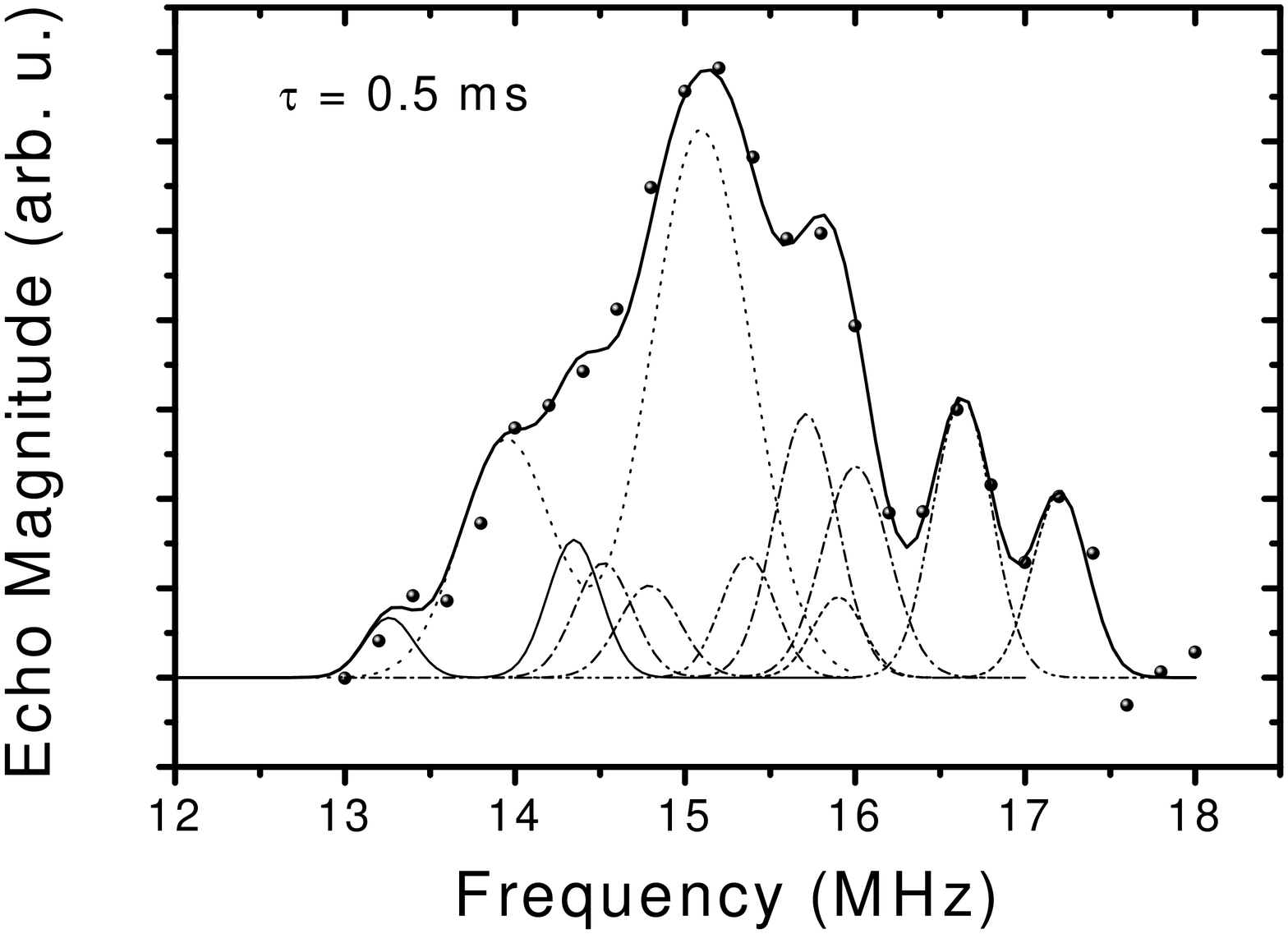}
\caption{\label{t2ox_60us}Measured $^{63/65}$Cu NQR spectra at
4.2~K under  various conditions for the originated sample. In the
upper picture the pulse spacing of the spin-echo sequence was
increased from 23 $\mu$s to 60 $\mu$s. In the lower picture the
frequency range to be measured was first saturated by a saturating
comp and after 0.5~ms the recovered magnetization was probed by a
spin-echo sequence. The relaxation times vary smoothly within the
spectrum.}
\end{figure}
freedom of the fits despite their large number of free parameters.
If one, for example, fits the lower end of the spectrum and
subsequently the upper end the resulting curve  has to fit the
middle part of the spectrum. Under this condition the resulting
fits are reliable if the measured spectra exhibits pronounced
peaks.
\begin{figure}[t]
\includegraphics [width=6cm, angle=0] {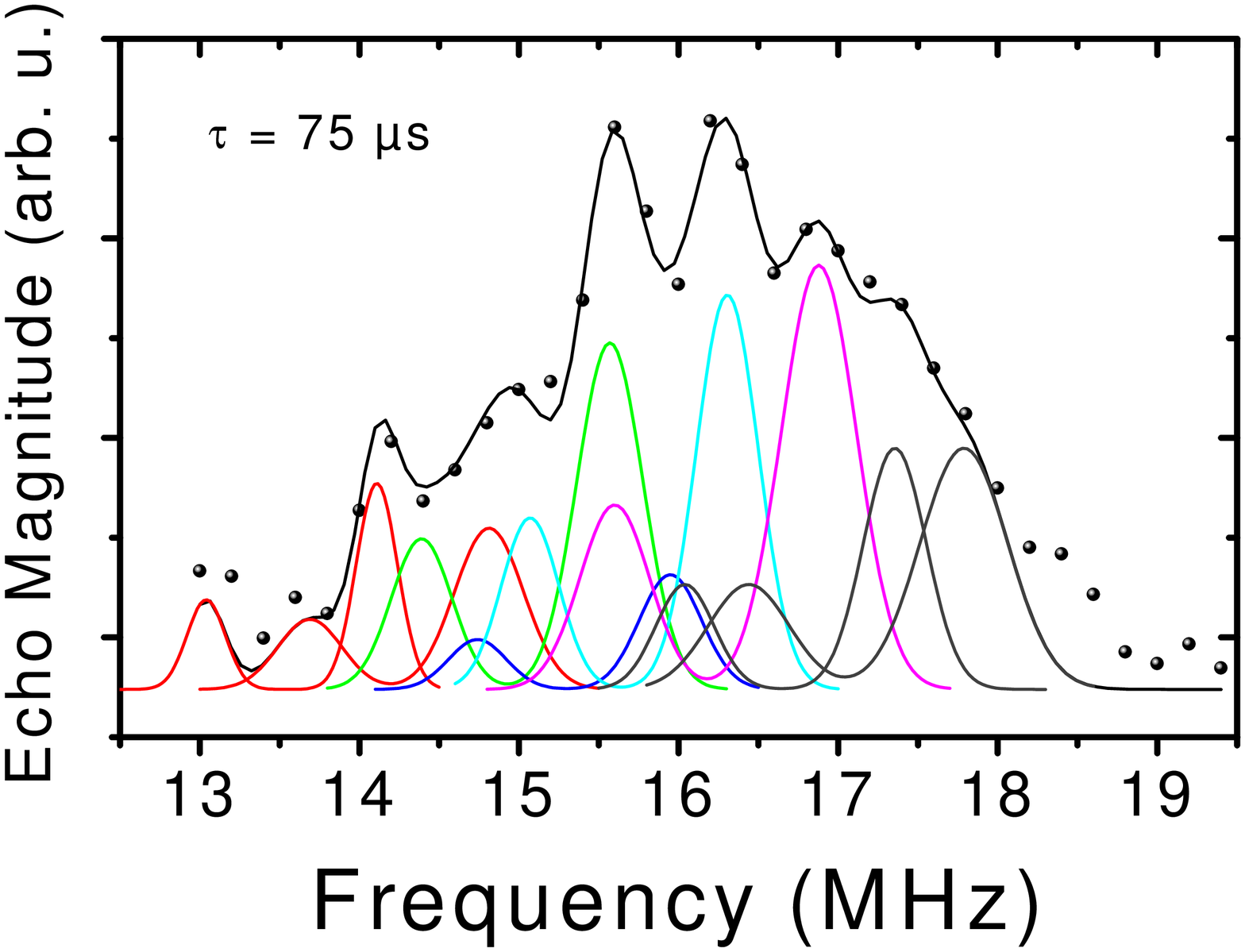}
\includegraphics [width=6cm, angle=0] {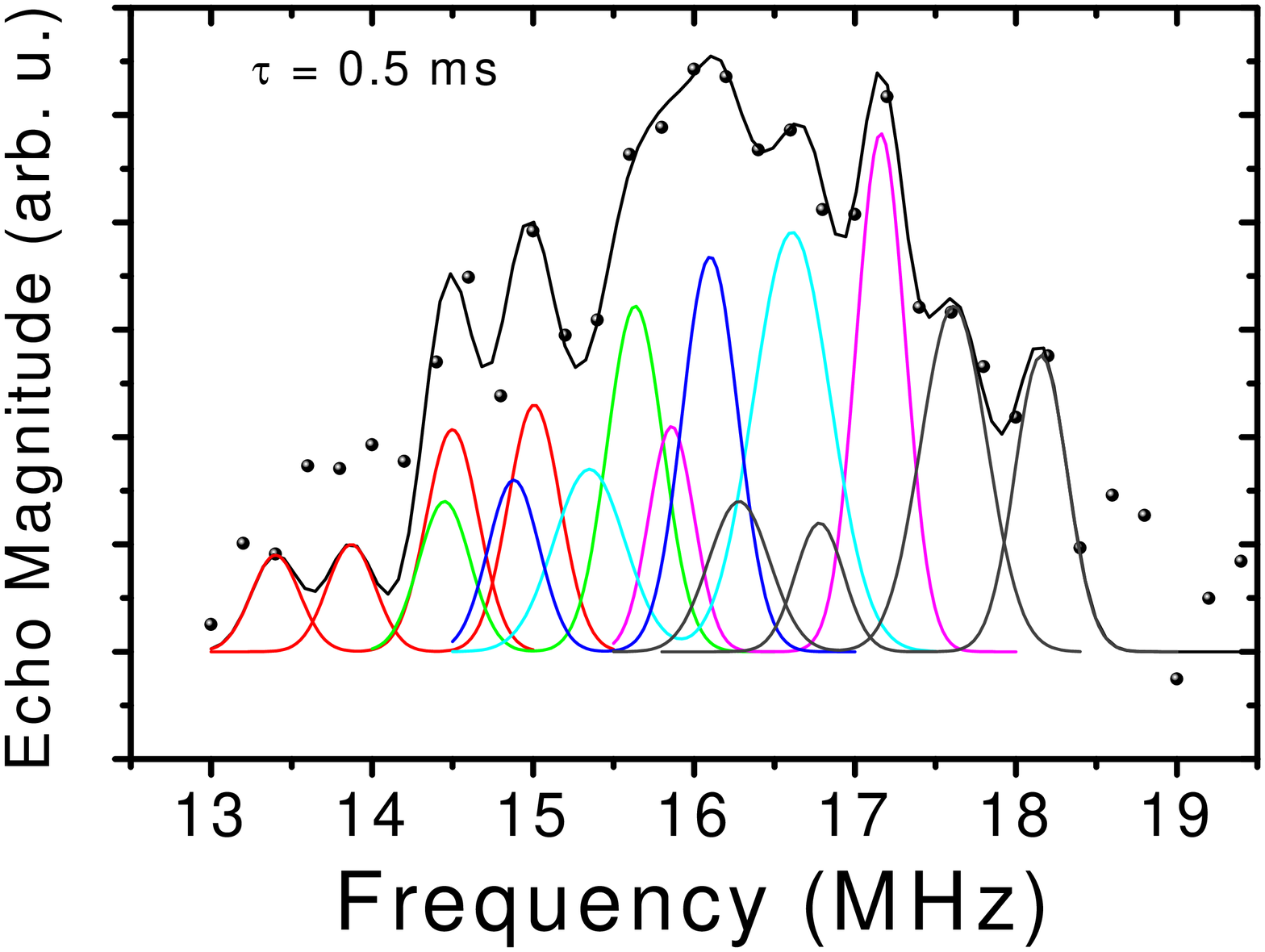}
\caption{Measured $^{63/65}$Cu NQR spectra at 4.2 K under  various
conditions for the fluorinated sample. In the upper picture the
pulse spacing of the spin-echo sequence was increased from 23
$\mu$s to 75 $\mu$s. In the lower picture the frequency range to
be measured was first saturated by a saturating comp and after
0.5~ms the recovered magnetization was probed by a spin-echo
sequence. The relaxation times vary smoothly within the spectrum.
} \label{t2f9_60us}
\end{figure}

The occurrence of additional lines within the spectrum is not an
evidence for a line splitting. These lines may arise from
impurities, Cu outside the CuO layers or Cu from distinct doped
phases. However, additional lines from impurities should be
clearly visible above $T_c$ within the spectrum and should
strongly differ from their relaxation times. The same can be
stated for lines arising from distinct doped phases.

\subsection{Spin-lattice relaxation}
\begin{figure}[h]
\includegraphics [width=6cm, angle=0] {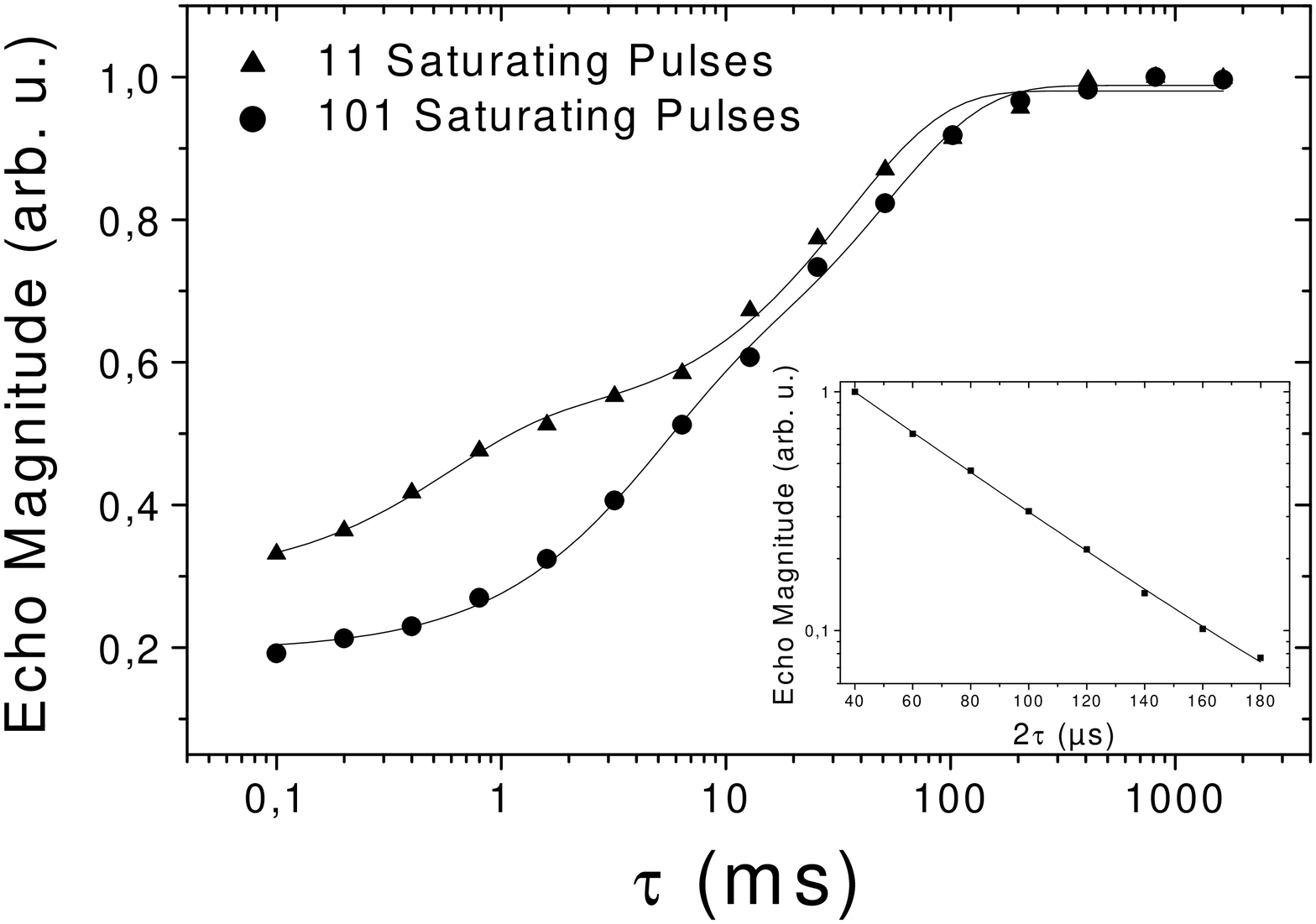}
\includegraphics [width=6cm, angle=0] {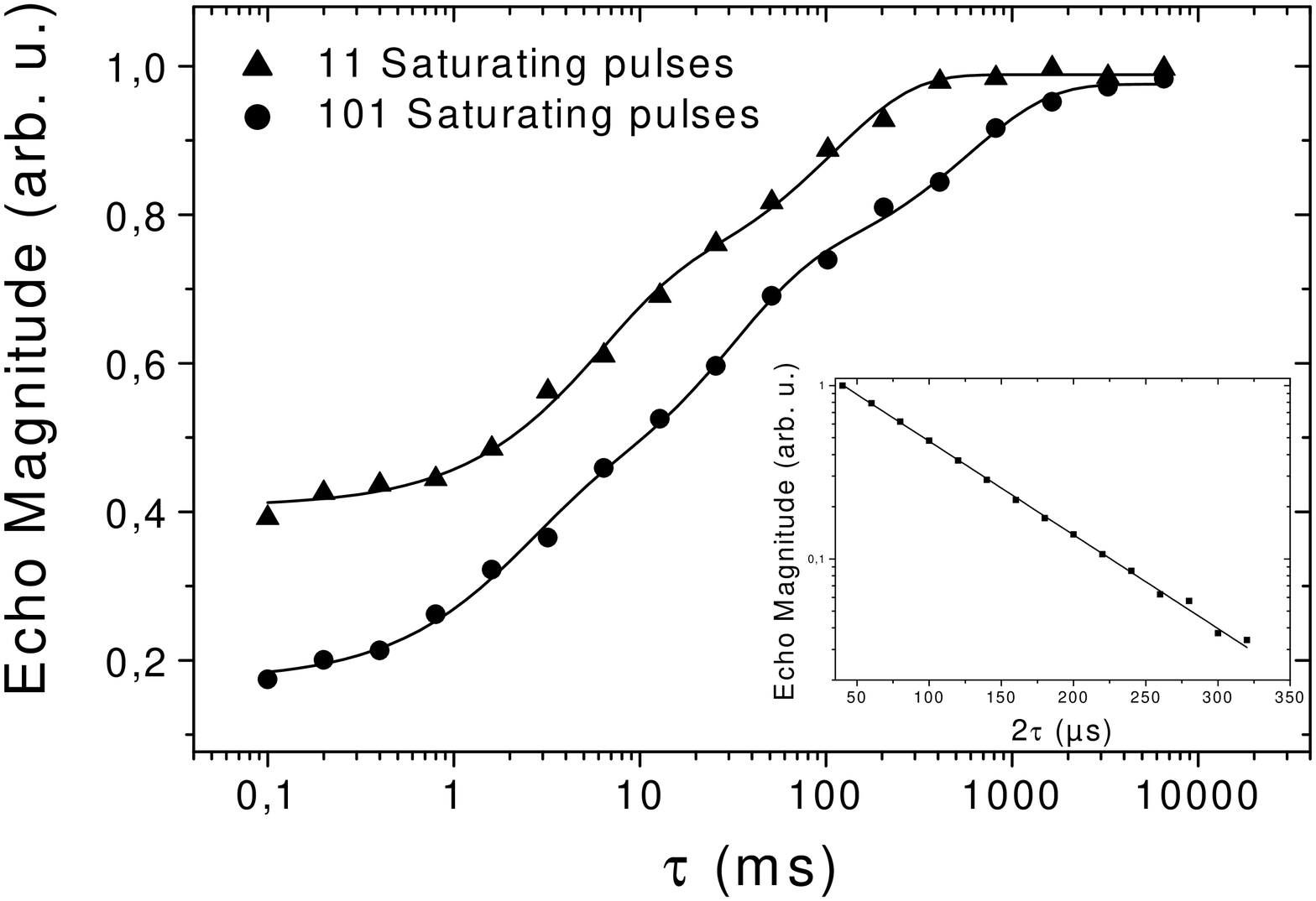}
\caption{ $^{63}$Cu spin-lattice relaxation curves at
$\nu$~=~16.6~MHz for the oxygenated (upper picture) and
fluorinated (lower picture) sample. The relaxation curve for 11
and 101 saturating pulses are denoted by squares and circles,
respectively. The solid curves in the upper picture denoting fits
using a two exponential function. The relaxation process in the
lower picture can be fitted with a two exponential function for 11
saturating pulses and a three exponential function for 101
saturating pulses. The $^{63}$Cu transverse relaxation is purely
single exponential for both samples (see insets).} \label{fig2}
\end{figure}

The spin-lattice relaxation time of Cu in superconducting
high-$T_c$s is extremely short and  driven by antiferromagnetic
(AF) fluctuation within the CuO layers \cite{suh}. The
spin-lattice relaxation times of Cu and Hg, in HgBaCuO for
example, differ by two orders of magnitudes \cite{suh}. From the
ratio $\gamma(Cu)/\gamma(Hg)$ of their gyromagnetic ratios one
would expect a ratio of the spin-lattice relaxation times of the
order of two. The difference is explained by the AF fluctuations
within the CuO layers strongly decreasing with distance. As
demonstrated by Figs. \ref{t2ox_60us} and \ref{t2f9_60us} the
relaxation times vary only smoothly within the spectra what can be
interpreted as a proof for Cu inside CuO layers as the origin of
the additional lines. Furthermore, these lines have to belong to
the same superconducting phase as proofed by the sharp
superconducting transition (Fig.~\ref{suscep}). From the doping
dependency of the Cu NQR frequency in the one layer superconductor
HgBaCaCuO\cite{gippius} one can estimate the difference in $T_c$
of phases possessing a Cu NQR frequency difference of 1 MHz to be
several tenth of K what is clearly not seen in Fig.~\ref{suscep}.
So far it is plausible that the additional are due to a line
splitting of the two NQR lines visible at 140~K. In order to
distinguish between the possible origins of the observed splitting
we present in Fig.~\ref{fig2} the $^{63}$Cu spin-lattice and
spin-spin relaxation curves measured at $\nu=$~16.6~MHz. Various
numbers of saturating pulses were applied according to the
dependency of the relaxation process illustrated in Sec.~\ref{spin
lattice}.
\begin{table}[t]
\caption{\label{tab:table1}$^{63}$Cu Spin-lattice relaxation times
of the oxygenated sample  $T^{(1)}_{1}$ and $T^{(2)}_{1}$ and the
according coefficients $a_1$ and $a_2$ of the two exponential fits
according to Fig.~\ref{fig2} for different numbers of saturating
pulses }
\begin{tabular}{l|c|r}
\label{tab1}
Pulses&$a_1$~~~~$T^{(1)}_{1}$(ms) &$a_2$~~~~$T^{(2)}_{1}$(ms)\\
\hline
11 &0.22(2)~~~0.6(2) &0.46(2)~~~35(4)\\
101 &0.44(3)~~~4.3(5) &0.35(3)~~~56(4) \\
\end{tabular}
\end{table}
\begin{table}[t]
\caption{\label{}$^{63}$Cu Spin-lattice relaxation times of the
fluorinated sample  $T^{(1)}_{1}$, $T^{(2)}_{1}$ and $T^{(3)}_{1}$
and the according coefficients $a_1$, $a_2$ and $a_3$ of the two
and three exponential fits according to Fig.~\ref{fig2} for
different numbers of saturating pulses }
\begin{tabular}{l|c|c|r}
\label{tab2}
Pulses&$a_1 T^{(1)}_{1}$(ms) &$a_2 T^{(2)}_{1}$(ms) &$a_3 T^{3}_1$(ms)\\
\hline 11  &0.28(2) 5.7(8) & 0.29(2) 108(15)
&--\\\hline
101 &0.23(3) 2.2(4) &0.32(2) 30(5) &0.26(2) 590(90)\\
\end{tabular}
\end{table}
As seen in Fig.~\ref{fig2} the spin-lattice relaxation follows a
two exponential behavior strongly depending on the number of
saturating pulses. The deduced relaxation times are shown in
Tab.~\ref{tab1}. It is remarkable that the deduced relaxation
times depend on the initial conditions. As discussed in
section\ref{spin lattice} this can be understood under the
assumption of three exponential relaxation behavior fitted with a
two exponential recovery law.  This assumption is supported by the
Cu spin-lattice relaxation in the fluorinated sample which is
shown in the lower picture of Fig.~(\ref{fig2}). Increasing the
number of saturating pulses reveals a third relaxation function.
Exactly such a behavior is expected from the relaxation theory and
equation~(\ref{satshort}) and (\ref{satlong}). The coefficient of
the component with the longer relaxation time increases with
increasing saturating time.
 Furthermore, the transverse relaxation is purely
single exponential. Thus, a formation of the internal magnetic
fields as an origin of the observed splitting can be concluded.

\section{Deduced magnetic fields}

A magnetic field present at the copper nuclei will split the two
line pairs observed at 140~K into four pairs of lines. However,
the spectra can not be fitted with four pairs of lines within the
whole temperature range. Instead, a reasonable agreement could be
achieved by using six pairs of lines. This can be understood by
the following assumption. If both outer CuO$_2$-layers would not
be equal below $T_c$, the NQR spectrum will consist of three pairs
of lines. In this case both outer CuO$_2$-layers differ with
respect to their EFG on the copper nuclei. If these three NQR line
pairs split due to the formation of magnetic moments in the
\begin{figure}[h]
\includegraphics [width=6.0cm,angle=0] {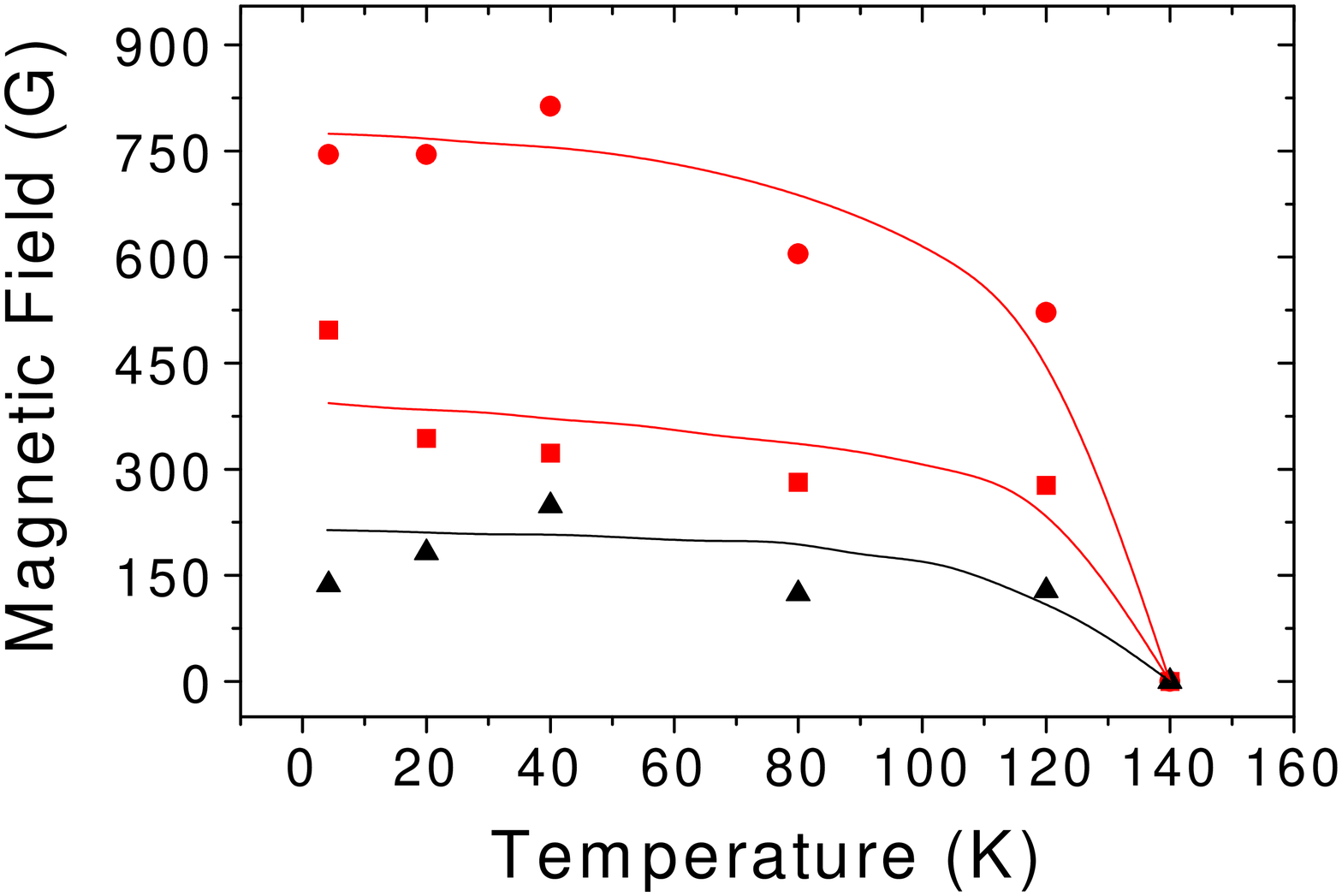}
\includegraphics [width=6.0cm, angle=0] {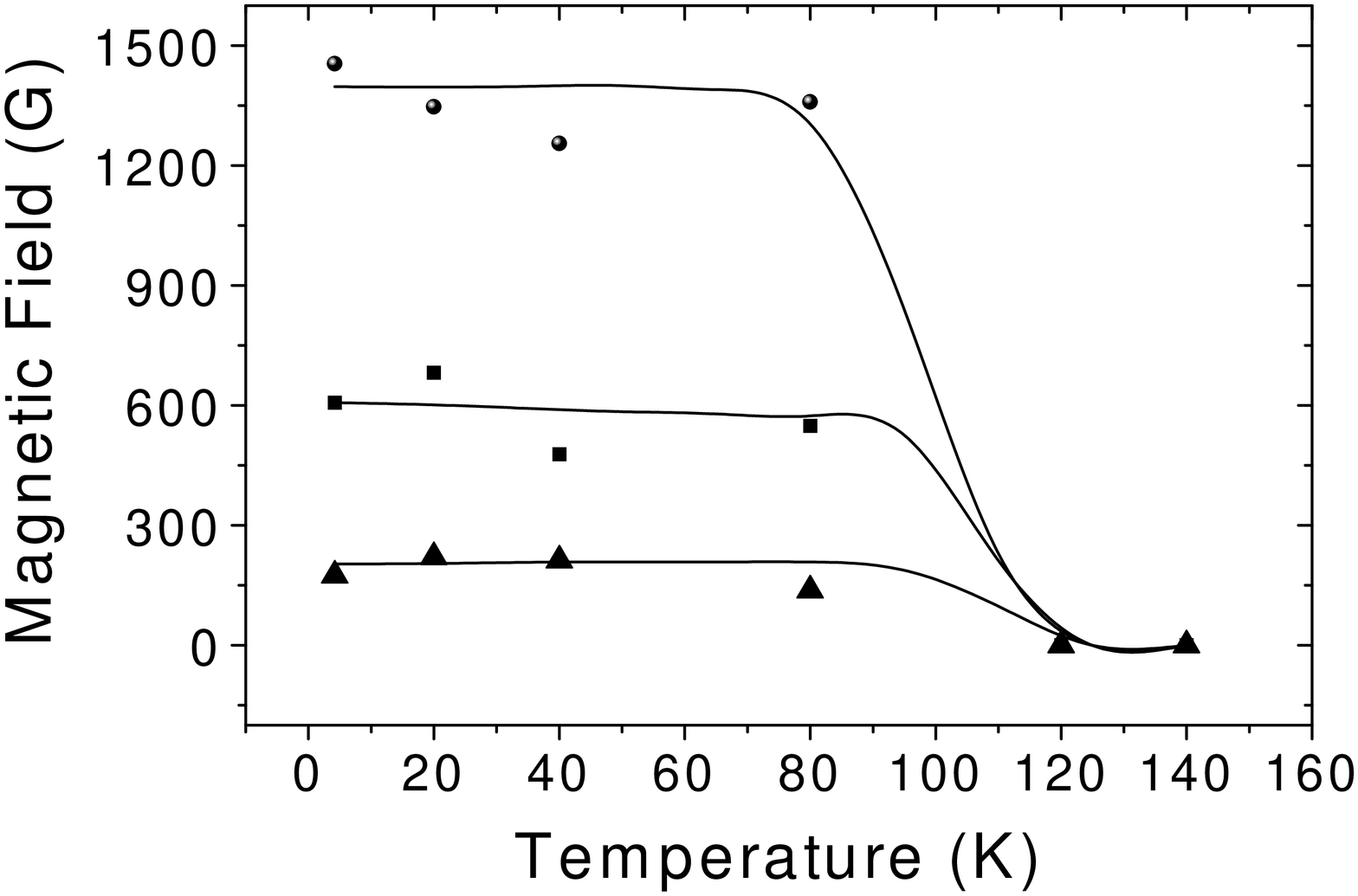}
\caption{\label{fig3} Temperature dependence of the extracted
magnetic field for the different CuO$_2$-layers (upper picture:
oxygenated sample, lower picture: fluorinated sample). Circles and
squares: outer layers, triangles: inner layer. The solid curves
are  guides to the eyes.  }
\end{figure}
CuO$_2$-layers, the spectrum will consist of six pairs of lines.
This is indeed observed. One may argue that an electrical
inequality is against the crystal symmetry because the inner CuO
layer is a symmetry layer. However, the symmetry is immediately
broken when a magnetic field appears in the inner layer with a
direction parallel to the c-axis. So far the rules of symmetry are
not violated. As discussed above the deduced three exponential
spin-lattice relaxation is as an indicator for exactly that field
direction. It can be stated that the interpretation of the data is
consistent.

Associating the line pairs to the CuO$_2$-layers has been done by
comparing the line intensities at 4.2~K under the condition that
the sum of the intensities of the outer CuO$_2$-layers lines have
to be twice of the the intensity of the inner CuO$_2$-layer.
Subsequently, the magnetic field has been deduced from the
equation $H=\Delta\nu/{2\gamma}$, with $\gamma$ as the
gyromagnetic ratio of the $^{63}$Cu nucleus and $\Delta\nu$ as the
frequency spacing of the $^{63}$Cu lines. The results are shown in
Fig.~\ref{fig3}. The values of the internal magnetic fields vary
from the inner to the outer CuO$_2$-layer and the maximum field is
of order of 700 Gauss for the originated sample and 1500 gauss for
the fluorinated sample .

Now we would like to discuss the possible origin of these internal magnetic
fields. We can exclude the presence of the water as a source of the
weak antiferromagnetism in the CuO$_2$-plane as was suggested
recently\cite{doogl}, since our measurements were performed on
the fresh samples. Furthermore, the values of the deduced magnetic fields
in our case is much larger than those measured in the water-inserted YBaCuO
compounds. Similarly, the formation of the time-reversal symmetry breaking
state \cite{varma} can be excluded since it would not give a magnetic
moment on the copper site.

Note that one of the most
probable scenarios for the microscopical origin of the observed magnetic
fields could be 
the formation of $id$-density wave \cite{laughlin} or the short-range
antiferromagnetic spin correlations remnant from the parent antiferromagnetic
compounds and suppressed by the doped carriers.
In the first
case, one expects that the $id$-density wave(DDW) state with the
d$_{x^2-y^2}$-wave
symmetry of the order parameter
($\eta_{\bf k} = i\frac{\eta_0}{2} (\cos k_x - \cos k_y)$)
is formed around the pseudogap formation temperature $T^*$ which for
optimally-doped cuprates is close to T$_c$.
Below $T^*$ the staggered orbital currents are
formed due to the imaginary character of the order parameter \cite{flux}.
Furthermore, these currents induce the formation of the internal
magnetic moments perpendicular to the CuO$_2$-layer and their magnitudes
are increasing upon decreasing temperature.
These magnetic moments also would
result in the splitting of the NQR lines due to the magnetic fields they
produce. In this scenario the value of the internal magnetic fields,
DDW gap, and pseudogap formation temperature will be ultimately
related to each other. In particular, using the relation obtained
earlier\cite{eremnew} between the orbital currents and the DDW
order parameter one gets $j_0 = \frac{4et}{\hbar J} \eta_0$ where $t=250$meV 
is a hopping integral between nearest neighbors in the CuO$_2$-plane and 
$J=120$meV 
is a corresponding superexchange interaction.
One can easily estimate the values of the
internal magnetic fields as $H_{int} = j(T=0)/ cr$
($r \approx 2~\AA$).
For example, taking the value of $\eta_0=30 meV$ one obtains 
$H_{int}$ of the order of 100G.
This value
qualitatively agrees with those we obtain from the splitting of NQR
lines at T$~= 4.2$~K for the inner CuO$_2$-layer. Note,
the corresponding internal magnetic moments are of the order
$l_c \approx 0.03 \mu_{B}$ that agrees with neutron scattering
experiments\cite{mook}. On the other hand, despite some
consistency we notice several problems.
For example, the magnitude
of the magnetic fields varies and enhances for the outer CuO$_2$-layers
in comparison to the inner CuO$_2$-layer. This seems to contradict
the $id$-density wave scenario, since it is expected that the
doping concentration of the
holes in the inner and the outer CuO$_2$-layers are different and the latter
are more doped. Then, the magnitude of the magnetic moments should be
smaller there. This is in contrast to our observation.
Therefore, further theoretical and experimental
studies are needed in order to identify the
microscopical origin of the internal magnetic fields.

\section{Summary} In summary, we perform copper NQR measurements
in the optimally-doped three-CuO$_2$-layer Hg-1223 high-temperature
cuprate superconductor.
We show that the $^{63}$Cu
NQR-lines in all three CuO$_2$-layers broaden below $T_c$ due to a
Zeeman splitting
originating from the internal magnetic fields (magnetic moments)
within the CuO$_2$-layers.
Analyzing the NQR spectra we obtain the values of the
internal magnetic fields within the CuO$_2$-layer. For the outer 
CuO$_2$-layers its values vary from $400$G and $1400$G, respectively,
while for the inner CuO$_2$-layer its value is approximately
$200$ G. Note, the formation of $id$-density wave or short-range
antiferromagnetic spin correlations remnant from the parent antiferromagnetic
compounds and suppressed by the doped carriers could be possible 
scenarios as a microscopic origin of the observed fields.

D.M. and I.E. are supported by the INTAS Grant No. 01-0654.


\end{document}